\documentclass[preprint,12pt,numbers]{elsarticle}

\usepackage{amssymb}
\usepackage{amsmath}
\usepackage[T2A,T1]{fontenc}
\usepackage[utf8]{inputenc}
\usepackage[russian,english]{babel}
\usepackage{graphicx}
\usepackage{placeins}
\usepackage{xurl}
\usepackage{hyperref}
\usepackage{seqsplit}
\usepackage{geometry}
\journal{Sensors and Actuators: A. Physical}

\begin{document}

\begin{frontmatter}



 \title{Study of Multilayer TiN$_x$/Ti/TiN$_x$ Films Formed In Situ for High-Efficiency Integrated Microheaters}


\author{Oleg~A.~Solovyev}
\author{Zakhar~S.~Enbaev}
\author{Alexander~R.~Shevchenko}
\author{Aleksander~M.~Mumlyakov}
\author{Pavel~A.~Berezhnoy}
\author{Mikael~O.~Geodakyan}
\author{Mikhail~V.~Tsvetkov}
\author{Michael~A.~Tarkhov \&}
\author{Igor~V.~Trofimov}
\affiliation{organization={Institute of Nanotechnology of Microelectronics of the Russian Academy of Sciences},
            addressline={}, 
            city={ Moscow},
            postcode={119334}, 
            country={Russia}}

\begin{abstract}
A technology for the in situ formation of a three-layer TiN$_x$/Ti/TiN$_x$ functional material within a single vacuum cycle is presented, together with the results of investigations of its structural and electrophysical properties. The maximum critical power density reached $\sim$140 MW/cm$^2$. Under continuous operation at 90\% of the critical power, the structures remained operational for 24 h. The characteristics of microbridges without and with a protective SiO$_x$ layer were compared. The presence of the coating modifies the electrothermal operating regime of the structures and enables estimated temperatures of $\sim$500–700 °C to be reached.
\end{abstract}

\begin{keyword}
Thin films, functional materials, TiN$_x$/Ti/TiN$_x$ multilayers,  integrated microheaters, gas sensors



\end{keyword}

\end{frontmatter}



\section{Introduction}
With the development of advanced microelectronic devices for gas sensing \cite{Bhattacharyya2014} and integrated photonics \cite{Lee2017}, miniature on-chip heating elements—microheaters—have emerged as one of the key components. Both fields require localized heating in a planar, CMOS-compatible configuration, incorporating dielectric membranes, protective coatings, and standard thin-film fabrication processes.

Catalytic gas sensors are used in industrial safety applications, including liquefied petroleum gas and natural gas monitoring and natural gas leak detection; environmental monitoring, particularly for methane as a greenhouse gas; mining applications for methane and CO detection; automotive exhaust monitoring systems; hydrogen safety in fuel-cell systems; and indoor air quality monitoring \cite{HarleyTrochimczyk2016}.

In a thermocatalytic gas detector, the microheater provides the thermal energy required to activate catalytic oxidation on the surface of the sensing element. The heat released during the reaction is converted into an electrical signal related to the concentration of the target gas \cite{Cattrall1997}.

The key parameters of gas sensors include power consumption, achievable temperature, process scalability, and response time. The development of autonomous wireless sensor nodes has established a power-consumption requirement of below 50 mW, and below 20 mW for the most efficient MEMS platforms \cite{HarleyTrochimczyk2016}. Current research is focused on the development of stable nanostructured catalytic layers that remain effective at reduced temperatures and are compatible with thick-film microelectromechanical systems technology \cite{Biro2020}.

For planar gas sensors, \textit{Roslyakov et al.} demonstrated that combining anodic oxidation of an aluminum substrate with magnetron sputtering enables the fabrication of heaters with operating temperatures above 500 °C, short response times, and a scalable technology suitable for semiconductor and thermocatalytic sensors \cite{Roslyakov2018}.

In integrated photonics, microheaters exploit the thermo-optic effect to modify the refractive index of the waveguide material \cite{Lee2017}. Their main applications include Mach–Zehnder interferometer-based switches, tuning of ring resonators, alignment of wavelength-division multiplexing channels, and stabilization of frequency combs. \textit{Masood et al.} demonstrated CMOS-compatible tungsten (W) microheaters fabricated using standard interconnect technology on an SOI platform. The power required for tuning over one free spectral range (FSR) was 35.2 mW, while the heating time from 10 to 90\% of the steady-state value was 2.8 $\mu$s \cite{Masood2014}.

\textit{Brian S. Lee et al.} developed and experimentally investigated a thermo-optic tuning scheme for suspended microresonators using wire bridges and platinum (Pt) microheaters. For a suspended microresonator on a silicon nitride platform, resonance tuning by 9.7 GHz was demonstrated at a heater power of 5.3 mW \cite{Lee2017}

\textit{Dmitriev et al.} demonstrated microheaters for planar photonics on a silicon nitride platform, fabricated from a three-layer ScN$_x$/Nb/ScN$_x$ stack. The heaters withstood current densities of $\sim$100 GA/m$^2$ and provided tuning over up to four FSRs ($\sim$600 GHz) at a power of $\sim$1 W \cite{dmitriev}.

In thermocatalytic gas sensors, the catalyst is in direct contact with the active heating element, which may accelerate its degradation. To mitigate the effect of the catalyst, CMOS-compatible protective dielectric coatings, particularly SiO$_x$ and Al$_2$O$_3$, are employed; similar layers are used for microheater passivation in planar photonics. \textit{Roslyakov et al.} used a thin Al$_2$O$_3$ film to protect the active element of a gas sensor \cite{Roslyakov2018}. However, non-equilibrium SiO$_x$ coatings formed by chemical vapor deposition may contain excess oxygen and participate in electromigration processes during microheater operation \cite{Filipovic2021}. \textit{Dmitriev et al.} described the degradation mechanism of SiO$_x$-coated heaters and showed that additional ScN$_x$ barrier layers above and below the main conductive layer of the microheater reduce the influence of the non-equilibrium dielectric matrix \cite{dmitriev}.

In this work, a three-layer TiN$_x$-Ti-TiN$_x$ stack formed in a single in situ vacuum process was developed and investigated as a functional material for the active element of a microheater.

The structural and electrophysical characteristics of the material were determined. It was shown that microbridges based on this material reach extremely high temperatures of up to $\sim$730 °C at a power of approximately 2 W and remain operational during prolonged operation at 90\% of the critical power.

\pagebreak
\section{Rationale for the Selection and Fabrication of the Functional Material for the Active Heating Element}
Thermocatalytic gas sensors and planar photonic devices employ similar functional materials and engineering approaches to microheater implementation, including membranes, isolation trenches, and buffer or protective coatings. These solutions enable a balance between power consumption, response time, temperature-field uniformity, and long-term reliability.

The material of the active element of a thermocatalytic detector should remain stable at temperatures of $\sim$300–600 °C and have a sufficiently high temperature coefficient of resistance (TCR), on the order of 3 \ensuremath{\times} 10$^-3$ °C$^-1$, to enable resistive temperature monitoring and provide adequate sensor sensitivity.
The most commonly used materials include Pt \cite{Houlet2006}, \cite{Roslyakov2018}, a-Si \cite{Brida1999}, \cite{HarleyTrochimczyk2016}, SiC \cite{HarleyTrochimczyk2016}, ITO \cite{Cakir2016}, and TiN \cite{Ito2018}. \textit{Houlet et al.} used 200 nm-thick platinum heaters with a titanium adhesion layer in thermoelectric hydrogen detectors \cite{Houlet2006}. \textit{Roslyakov et al.} investigated Pt microheaters with functional-layer thicknesses of 300 and 800 nm and compared their operating parameters \cite{Roslyakov2018}. \textit{Brida S. et al.} used a 450 nm-thick polysilicon layer deposited by LPCVD as the heating layer of a MOS detector. In the membrane configuration, a temperature of 500 °C was achieved at a power consumption of 30 mW \cite{Brida1999}. \textit{Harley-Trochimczyk et al.} compared gas detectors incorporating polysilicon and 130 nm-thick 3C-SiC microheaters. In the membrane configuration, the SiC microheater exhibited more stable resistance than the polysilicon microheater during continuous heating at 500 °C and thermal cycling in the range of 500–650 °C. Using a thermally stable catalytic material—platinum nanoparticle-loaded boron nitride aerogel—reproducible propane detection over 18 h was demonstrated, with response and recovery times of less than 1 s \cite{HarleyTrochimczyk2016}.

Indium tin oxide (ITO) is an alternative material for planar heaters. \textit{M. Cihan Çakır et al.} integrated a 140 nm-thick ITO resistor beneath the SnO$_2$ sensing layer of a membrane-based gas sensor. An operating temperature of 420 °C was achieved at a power consumption of 120 mW; the temperature nonuniformity was $\sim$17 °C at an average heater temperature of 290 °C. The rise and fall times of the thermal response were 40 and 20 ms, respectively \cite{Cakir2016}.

Titanium nitride (TiN) is used in high-temperature gas detectors, including MEMS devices, owing to its electrical conductivity, thermal stability, and compatibility with thin-film fabrication processes. \textit{Ito et al.} fabricated TiN microheaters by photolithography using a sacrificial titanium layer followed by reactive magnetron sputtering. The maximum heating temperature reached $\sim$800 °C, while the dissipated power at 400 °C was $\sim$170 mW \cite{Ito2018}.

The search for low-cost and technologically viable materials for thermocatalytic gas sensing is ongoing. One approach is to eliminate precious metals and complex membrane structures in favor of scalable parallel fabrication. \textit{Dmitriev et al.} investigated microheater degradation at current densities of up to 100 GA/m$^2$ and proposed barrier layers to isolate the main conductive layer from the non-equilibrium oxide \cite{dmitriev}.

\begin{figure}[!htbp]
    \centering
    \includegraphics[width=\linewidth]{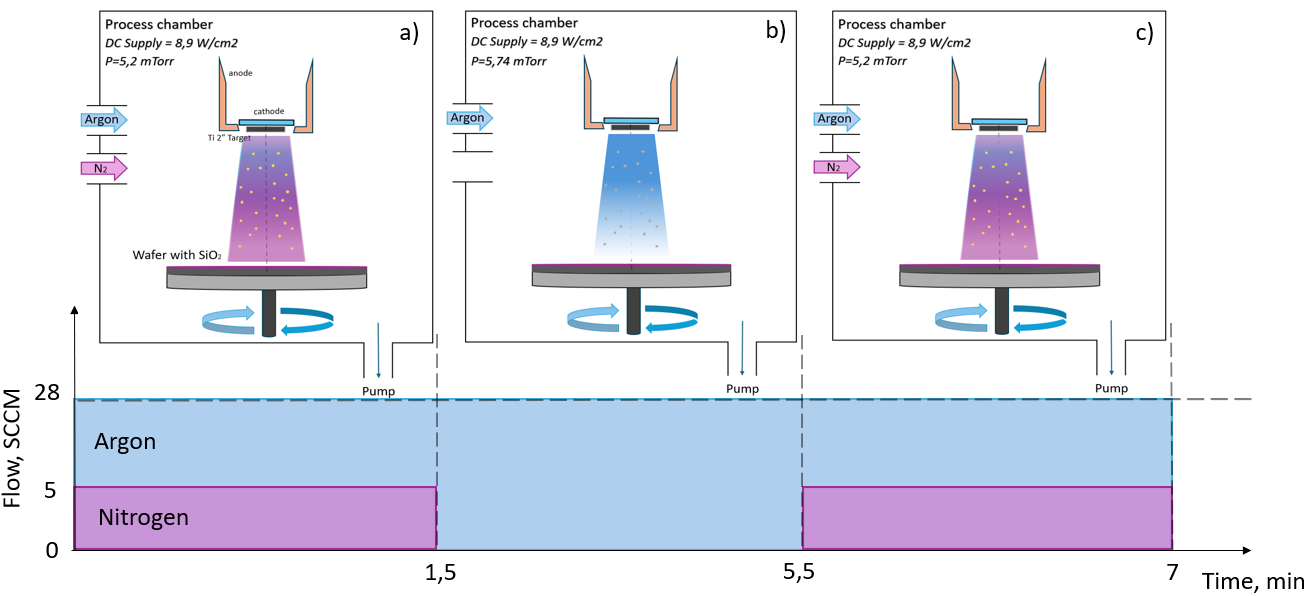}
    \caption{Schematic of the TiN$_x$/Ti/TiN$_x$ stack formation: (a) reactive deposition of the bottom TiN$_x$ layer in an Ar/N$_2$ mixture; (b) deposition of the Ti layer in Ar without reactive-gas supply; (c) reactive deposition of the top TiN$_x$ layer in an Ar/N$_2$ mixture.}.
    \label{fig:dst}
\end{figure}

In this work, the functional layer of the planar microheater consisted of a three-layer TiN$_x$/Ti/TiN$_x$ stack. Magnetron sputtering with controlled reactive-gas supply enables sequential deposition of layers with different compositions within a single vacuum cycle, thereby minimizing contamination at the interfaces between the layers. The stack was formed on 100 mm-diameter p-type silicon wafers with (100) orientation. After cleaning in acidic and alkaline solutions using a surfactant, Caro’s mixture, and diluted HF, a $\sim$4 $\mu$m-thick thermal SiO$_x$ layer was grown on the substrates. TiN$_x$/Ti/TiN$_x$ deposition was carried out using a Triple Target Sputter Coater DST3 equipped with a single two-inch titanium target. Argon (Ar) was used as the working gas; during the TiN$_x$ deposition stages, nitrogen (N$_2$) was additionally introduced into the chamber. All three layers were deposited sequentially without breaking vacuum (Figure \ref{fig:dst}).

\FloatBarrier

In the first stage, the bottom TiN$_x$ barrier layer was formed at an Ar:N$_2$ flow ratio of 5.6:1 and a pressure of 5.2 mTorr (Figure \ref{fig:dst}(a)). In the second stage, the N$_2$ supply was stopped and Ti was deposited in Ar at a pressure of 5.74 mTorr (Figure \ref{fig:dst}(b)). The N$_2$ supply was then restored to the initial Ar:N$_2$ ratio of 5.6:1, and the top TiN$_x$ barrier layer was formed at a pressure of 5.2 mTorr (Figure \ref{fig:dst}(c)). At all stages, the magnetron power was 180 W, corresponding to a target power density of 9 W/cm$^2$. The substrate holder temperature was 23 °C

\subsection{Structural Characterization of the Functional Material}

\begin{figure}[ht]
    \centering
    \includegraphics[width=\linewidth]{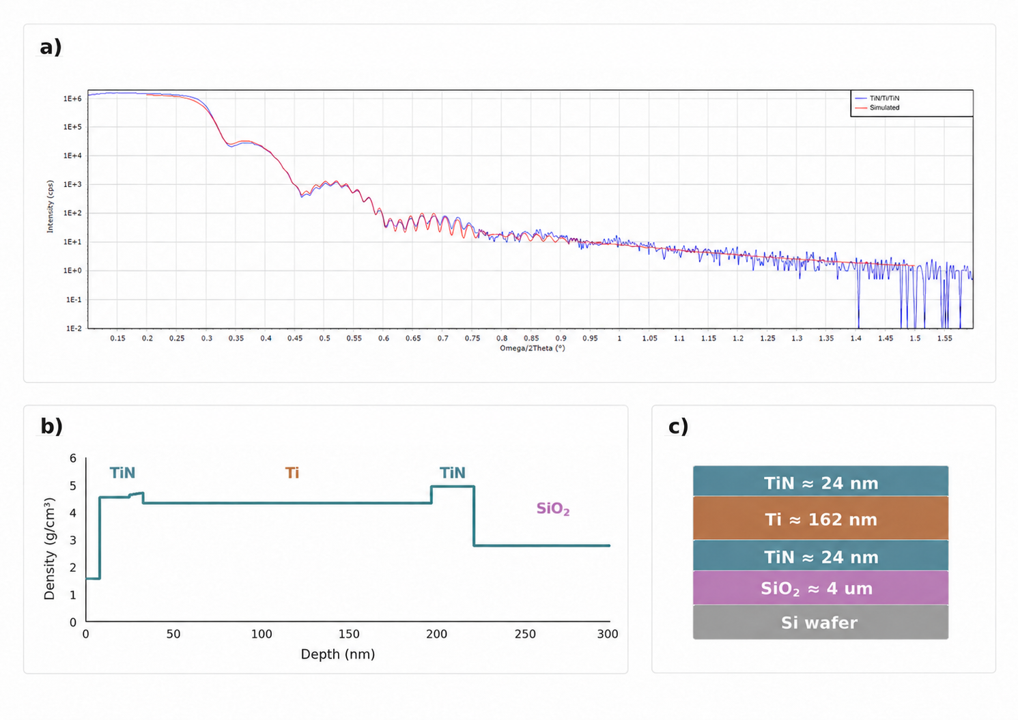}
    \caption{(a) Experimental XRR curve of the multilayer film; (b) depth-dependent density profile obtained by fitting the XRR curve; (c) schematic representation of the TiN$_x$/Ti/TiN$_x$ layers with estimated thicknesses.}
    \label{fig:xrr}
\end{figure}

The layer thicknesses were determined by X-ray reflectometry (XRR) using a PANalytical Empyrean diffractometer. The reflectivity curves were recorded in the $\omega$–2$\theta$ geometry using Cu K$\alpha$ radiation ($\lambda$ = 1.540605 Å) at an accelerating voltage of 45 kV and a current of 40 mA. The total film thickness was $\sim$220 nm. The absence of pronounced oscillations in the XRR curve limits the reliability of determining the thicknesses of the individual layers, since the oscillation period and amplitude contain information on the layer thicknesses and electron-density contrast. As the total film thickness increases, the oscillation period decreases, and at limited angular resolution the oscillations may become indistinguishable. Surface and interfacial roughness also accelerates the damping of specular oscillations due to diffuse scattering. Therefore, the thicknesses of the individual layers shown in Figure \ref{fig:xrr} should be regarded as estimates.

\FloatBarrier

\begin{figure}[!htbp]
    \centering
    \includegraphics[width=\linewidth]{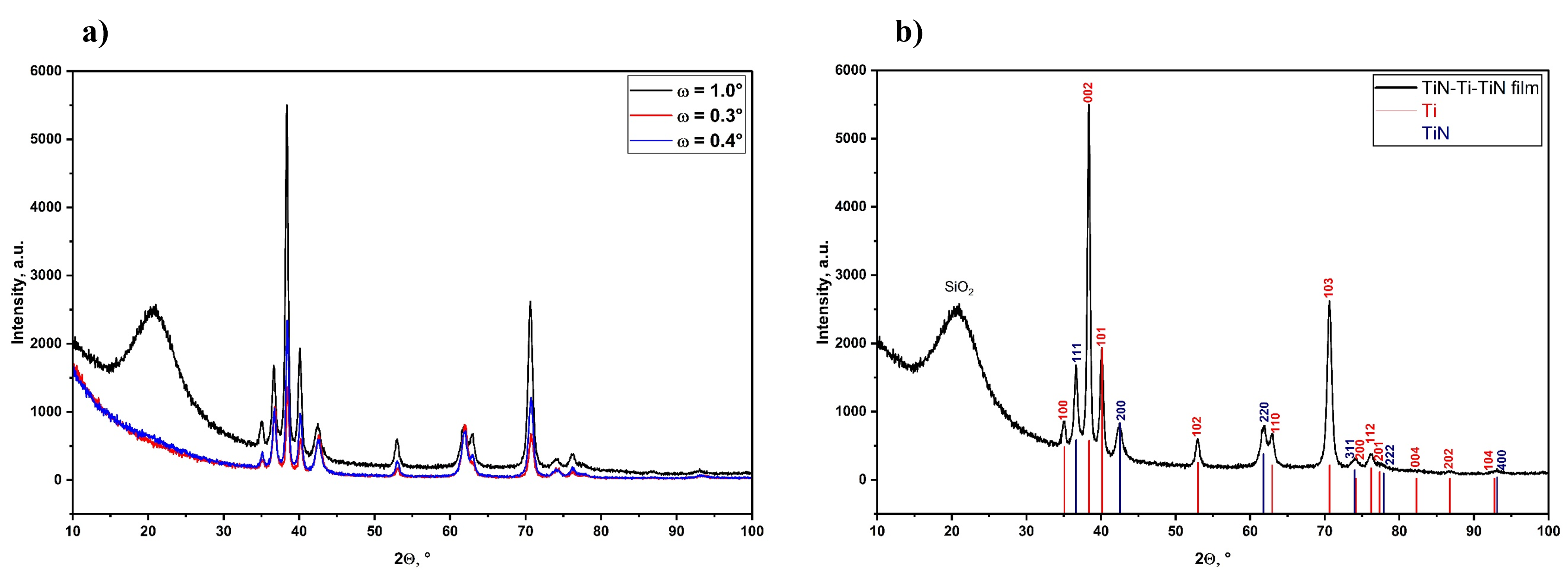}
    \caption{(a) GIXRD patterns of the TiN$_x$/Ti/TiN$_x$ sample at $\omega$ = 0.3°, 0.4°, and 1.0°; (b) phase identification of the diffraction pattern at $\omega$ = 1.0°. Blue tick marks indicate the positions of reflections corresponding to the cubic TiN$_x$ phase, while red tick marks indicate those of the hexagonal Ti phase. }
    \label{fig:xrd}
\end{figure}
\FloatBarrier

The crystal structure of the TiN$_x$/Ti/TiN$_x$ stack was investigated by grazing-incidence X-ray diffraction (GIXRD) at incidence angles of $\omega$ = 0.3°, 0.4°, and 1.0° (Figure \ref{fig:xrd}). The positions of the diffraction peaks correspond to the cubic titanium nitride phase and the hexagonal metallic titanium phase.

The broad feature in the 2$\theta$ = 15–25° range is predominantly associated with the amorphous SiO$_x$ layer of the substrate. The titanium layer exhibits a pronounced preferred orientation along the (002) and (103) directions, whereas the TiN$_x$ (111) texture is less pronounced. The Ti(002)/Ti(103) intensity ratio remains nearly constant with changing incidence angle (\ensuremath{\approx}2.0 for $\omega$ = 0.3–1.0°), indicating a similar distribution of crystallographic orientations throughout the probed depth of the titanium layer.

\begin{figure}[!htbp]
    \centering
    \includegraphics[width=\linewidth]{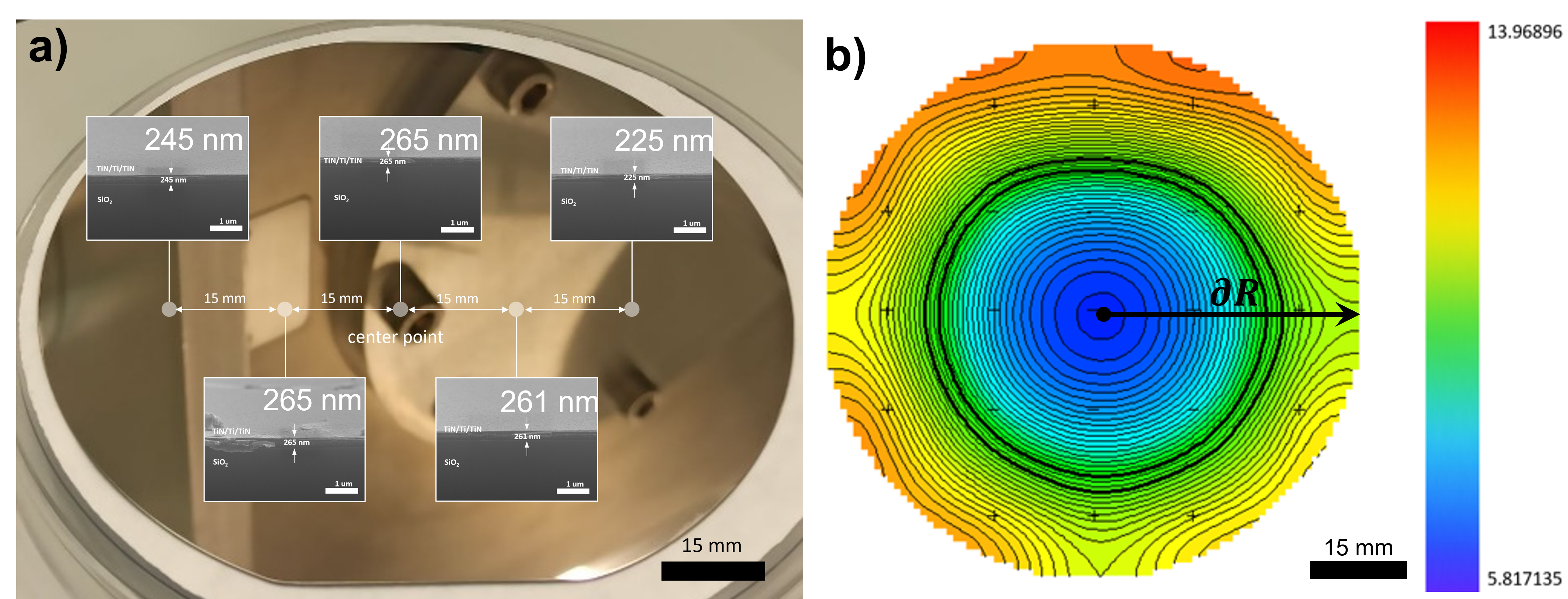}
    \caption{(a) Locations of SEM measurements of the film thickness on a 100 mm-diameter wafer; (b) sheet-resistance distribution map, $R_s$ (\ensuremath{\Omega}/\ensuremath{\square}).}
    \label{fig:wafer}
\end{figure}

Additionally, the stack thickness was measured on a cross-sectional cleave using scanning electron microscopy (SEM, Thermo Fisher). Measurements were performed at five locations: at the center of the wafer and at distances of 15 and 30 mm from the center in two opposite directions (Figure \ref{fig:wafer}a). Film uniformity was evaluated from a sheet-resistance map, $R_s$, measured by the four-point probe method using a 4Dimensions 280SI system (Figure \ref{fig:wafer}b).

\FloatBarrier

The average sheet resistance was \ensuremath{\approx}10 \ensuremath{\Omega}/\ensuremath{\square}. According to the $R_s$ map, the within-wafer nonuniformity reached \ensuremath{\approx}27\% (Figure \ref{fig:wafer}b). Assuming a spatially uniform electrical resistivity, such a variation in $R_s$ corresponds to a comparable nonuniformity in the thickness of the functional layer. The main reason may be the mismatch between the target diameter of 50.8 mm (2 inches) and the substrate diameter of 100 mm (4 inches). The nonuniformity can potentially be reduced by increasing the target diameter and optimizing the deposition geometry.

\section{Experimental Section}
\subsection{Experimental Determination of the Temperature Coefficient of Resistance}

The temperature coefficient of resistance (TCR) determines the sensitivity of resistive temperature monitoring of a microheater. To construct a thermal model, the $R(T)$ dependence of the specific material must be known over the operating temperature range. A high TCR increases the sensitivity of electrical thermometry; however, the accuracy of temperature estimation also depends on heating uniformity and the structural stability of the material \cite{Das2014TCR}. In the linear approximation, the resistance dependence is expressed as:

\begin{equation}
R_T = R_0 \left(1 + \alpha \left(T - T_0\right)\right)
\label{eq:tcr}
\end{equation}

where R$_T$ and R$_0$ are the electrical resistance values at temperatures T and T$_0$, respectively, and $\alpha$ is the TCR, expressed in K$^{-1}$ (numerically equivalent to °C$^{-1}$ for temperature increments). The reference temperature T$_0$ is selected within the calibration range; in the present work, the initial reference temperature was room temperature, 23 °C. From Equation (\ref{eq:tcr}), it follows that:

\begin{equation}
\alpha =
\frac{R_T - R_0}{R_0\left(T - T_0\right)}
=
\frac{1}{R_0}\frac{\partial R}{\partial T}
\label{eq:tcr_general}
\end{equation}

The TCR of the three-layer TiN$_x$/Ti/TiN$_x$ material was determined using dies without patterned topological features, selected at several positions along the diameter of the 100 mm wafer. The resistance was measured at fixed temperatures from 23 to 250 °C with a temperature step of $\partial T$ = 50 °C.

At each temperature step, the sample was held for 5 min to allow the temperature to stabilize. The measurements were performed on a temperature-controlled Sawatec contact hot plate, and the resistance was recorded using a Fluke 87V multimeter.

\begin{figure}[!htbp]
    \centering
    \includegraphics[width=0.5\linewidth]{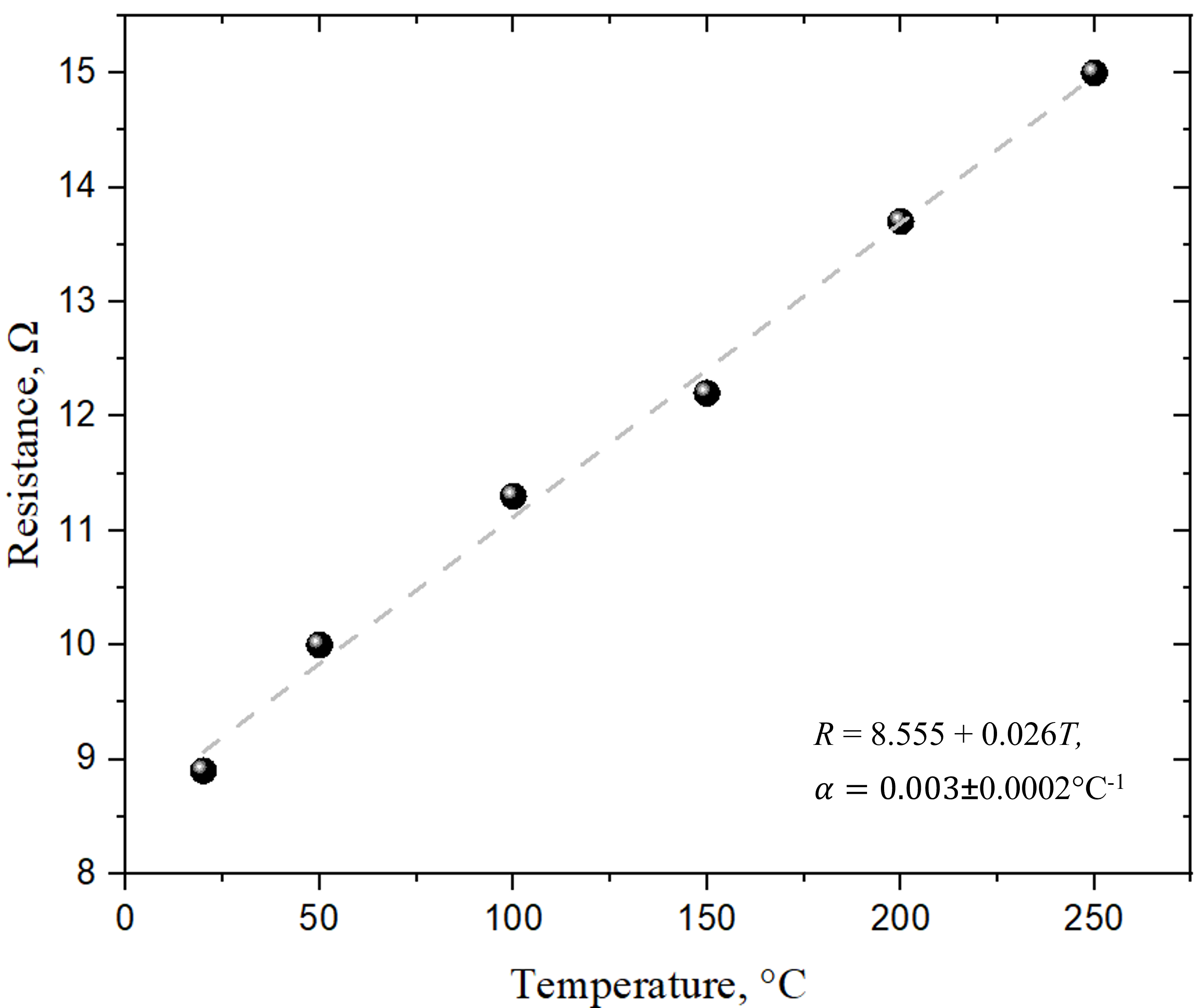}
    \caption{Temperature dependence of the resistance of TiN$_x$/Ti/TiN$_x$ and its linear fit; average TCR $\alpha = (3.0 \pm 0.2) \times 10^{-3}$ °C$^{-1}$.}
    \label{fig:tkc}
\end{figure}

The experimental $R(T)$ dependence and the calculated average TCR are presented in Figure \ref{fig:tkc}.

\FloatBarrier

\subsection{Formation of Topological Features}

\begin{figure}[h]
    \centering
    \includegraphics[width=\linewidth]{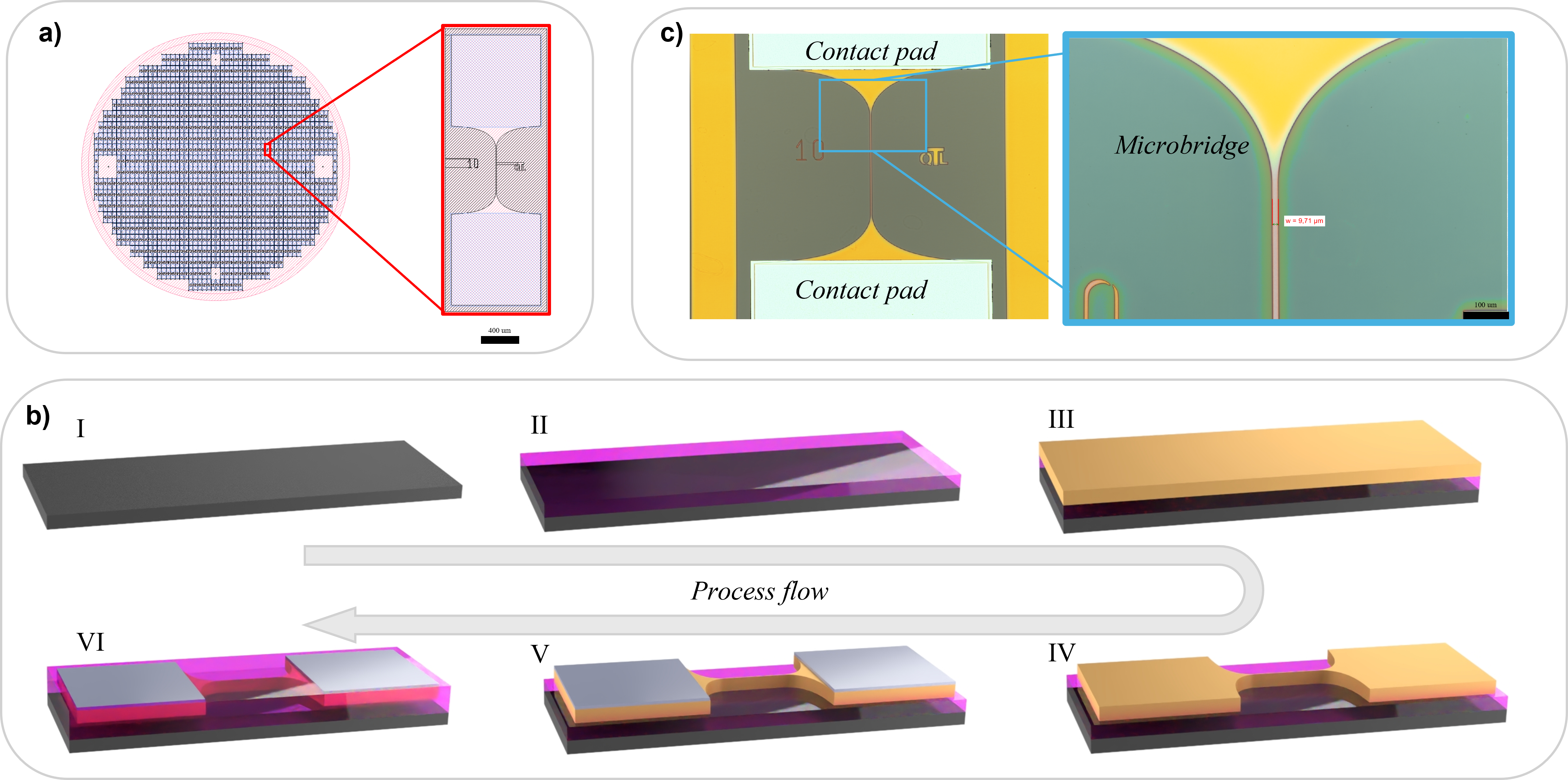}
    \caption{(a) Microbridge layout on a 100 mm-diameter wafer; (b) fabrication process flow: I — silicon wafer preparation; II — SiO$_x$ formation; III — TiN$_x$/Ti/TiN$_x$ deposition; IV — microbridge formation; V — formation of Al contact pads; VI — deposition of the protective SiO$_x$ layer and opening of the contact pads; (c) optical image of the fabricated microbridge.}
    \label{fig:bridges}
\end{figure}

To investigate the electrophysical characteristics of TiN$_x$/Ti/TiN$_x$, microbridge structures were fabricated on a 100 mm-diameter wafer (Figure \ref{fig:bridges}a). The microbridge width was varied from 1 to 10 $\mu$m in 1 $\mu$m increments, while the length was kept constant at 650 $\mu$m. The gradual widening of the conductor toward the contact pads reduced local current-density crowding and the probability of premature failure.

The fabrication process flow for the topological structures is shown in Figure \ref{fig:bridges}b and includes CMOS-compatible steps for forming the conductive and protective layers.

After substrate preparation and TiN$_x$/Ti/TiN$_x$ deposition, the microbridge topology was patterned by maskless laser direct imaging (LDI).

A 700 nm-thick MicroChemicals AZ® 1505 photoresist layer was used as the etch mask. Dry plasma etching by inductively coupled plasma reactive ion etching (ICP-RIE) was performed in a Cl$_2$:BCl$_3$:HBr gas mixture with a flow ratio of 2:1:2.

For electrical contacts, aluminum pads were fabricated using a lift-off process. The pattern was formed in an \ensuremath{\approx}6 $\mu$m-thick spray-coated \foreignlanguage{russian}{ФРАСТ ФП-РН-7С} photoresist, followed by ion activation of the surface, magnetron sputtering of a 400 nm-thick Al layer, and removal of the resist mask in organic solvents.

Some of the structures were coated with a 1 $\mu$m-thick protective SiO$_x$ layer deposited by PECVD using TEOS as the precursor. Above the contact pads, the coating was locally removed after maskless laser lithography by plasma etching in a fluorocarbon C$_4$F$_8$ plasma. The fabricated structure is shown in Figure \ref{fig:bridges}c.

\FloatBarrier

\subsection{Experimental Investigation of Critical Power Parameters}

To determine the critical power and current-density values corresponding to microbridge failure, the current–voltage characteristics of the structures were measured. A Karl Suss PM5 probe station and a Rohde\&Schwarz HMP4040 programmable power supply were used.

The wafer containing the structures was placed on the probe-station stage. Tungsten probes were brought into contact with the contact pads, after which automated measurements were performed in LabVIEW. Switching to the next structure was carried out manually.

The current was increased stepwise from 0 in increments of 0.1 mA while the voltage was recorded simultaneously, until either the source limit was reached or the microbridge failed. The power, $P = UI$, and the differential resistance, $dU/dI$, were calculated from the resulting $U(I)$ dependences.

\begin{figure}
    \centering
    \includegraphics[width=\linewidth]{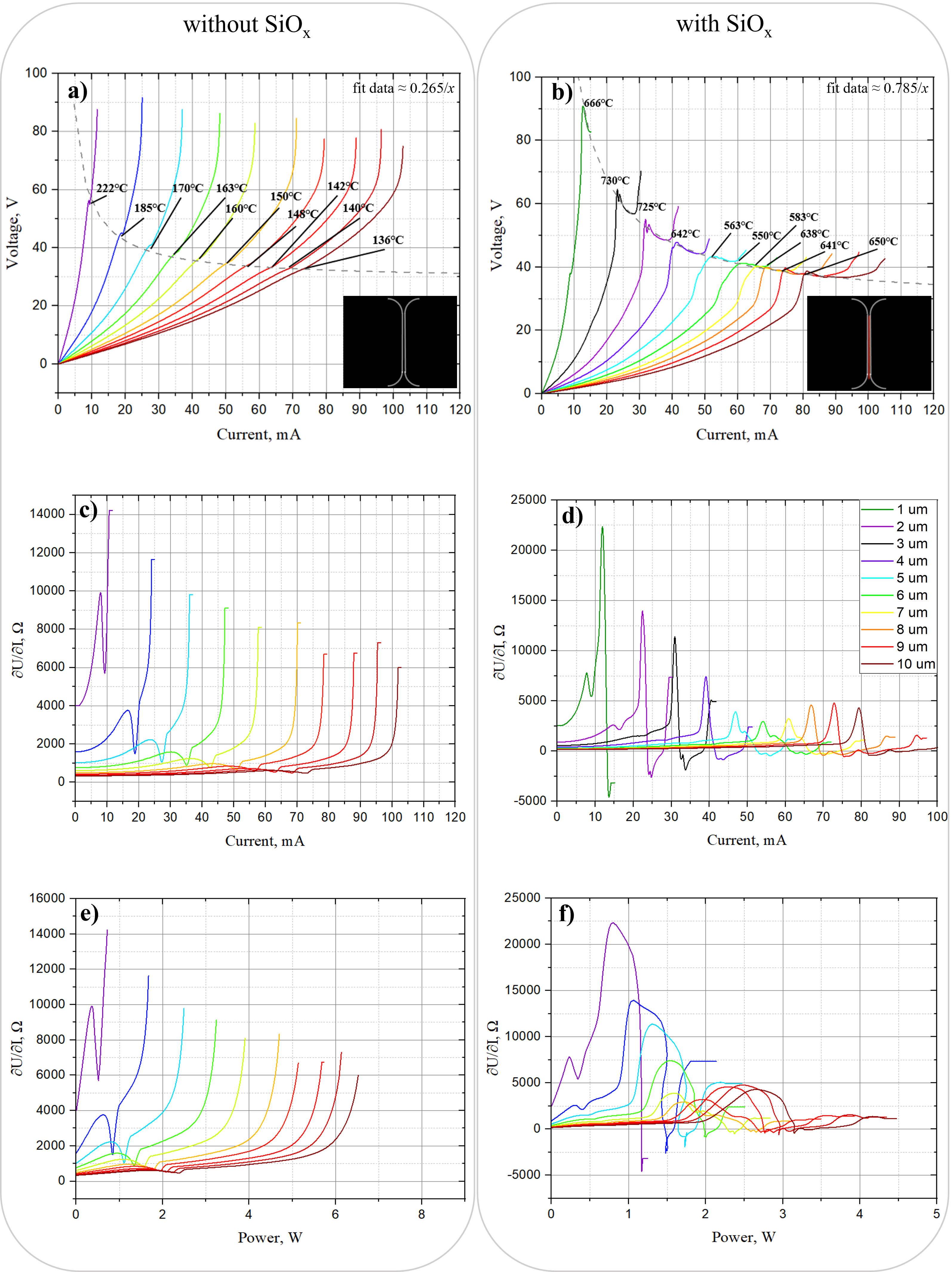}
    \caption{Electrophysical characteristics of TiN$_x$/Ti/TiN$_x$ microbridges with widths of 1–10 $\mu$m. (a, b) Current–voltage characteristics and optical images of the structures without and with a protective SiO$_x$ layer, respectively; the numbers next to the curves indicate temperatures estimated from the $R(T)$ dependence. (c, d) Differential resistance $dU/dI$ as a function of current. (e, f) $dU/dI$ as a function of dissipated power, $P = UI$. The left column corresponds to structures without SiO$_x$, and the right column to structures with SiO$_x$. Dashed lines in panels (a, b) show the envelope fits presented in the plots.}
    \label{fig:vacpw}
\end{figure}

The sweep rate, power-supply configuration, and ambient conditions were kept constant. Microbridges with widths ranging from 1 to 10 $\mu$m were investigated both without a coating and with a protective SiO$_x$ layer. The samples were selected within 30 mm of the wafer center, where the thickness variation was minimal (Figure \ref{fig:wafer}). The current–voltage characteristics were used to determine the boundary of stable electrothermal operation, the critical current and power values, and to estimate the temperature using the previously measured $R(T)$ dependence. The results are presented in Figure \ref{fig:vacpw}.

\pagebreak

\FloatBarrier

\pagebreak

\section{Results and Discussion}

Figure \ref{fig:vacpw} compares the electrothermal characteristics of microbridges with widths of 1–10 $\mu$m, both without a protective coating and with a 1 $\mu$m-thick SiO$_x$ coating. In the initial regions of the I–V characteristics, the voltage increases superlinearly with current due to Joule heating and the positive TCR of the material. As the microbridge width decreases, the nonlinear regime is reached at a lower current, corresponding to the increase in current density caused by the reduced cross-sectional area. For the uncoated structures, the $U(I)$ dependences remain predominantly monotonic up to failure (Figure \ref{fig:vacpw}a), whereas the coated samples exhibit pronounced voltage maxima followed by descending branches (Figure \ref{fig:vacpw}b).

The temperatures calculated from the resistance change are approximately 136–222 °C for the uncoated microbridges and 550–730 °C for the SiO$_x$-coated structures. Therefore, it is more appropriate to describe the difference as an increase of several hundred degrees rather than as a multiplicative increase in temperature on the Celsius scale. The visible glow of the coated microbridge in the optical inset (Figure \ref{fig:vacpw}b) is consistent with the transition to a high-temperature regime, whereas no visible glow is observed for the uncoated sample (Figure \ref{fig:vacpw}a). This difference may be associated with changes in heat-dissipation conditions after SiO$_x$ deposition; however, attributing it solely to suppressed convection would be insufficient, since the coating also modifies the effective thermal resistance of the system, interfacial heat transfer, and the surface emissivity. Distinguishing between these contributions requires a coupled thermal model or an independent temperature measurement.

The temperature was estimated from the room-temperature resistance and the measured TCR. However, the $R(T)$ calibration was performed only over the temperature range of 23–250 °C, whereas the maximum values in Figure \ref{fig:vacpw}b reach 730 °C. Consequently, the high-temperature values were obtained by extrapolating the linear dependence and should be regarded as estimates; their accuracy may decrease due to nonlinearity of $R(T)$, annealing, changes in stoichiometry, or oxidation of the film. The dashed envelopes in Figures \ref{fig:vacpw}a and \ref{fig:vacpw}b were fitted with a dependence of the form $U\propto I^{-1}$. For such a dependence, the product $UI$ is approximately constant; the increase in the coefficient indicated in the plots from \ensuremath{\approx}0.265 to \ensuremath{\approx}0.785 qualitatively reflects the shift in the characteristic electrothermal regime after SiO$_x$ deposition.

Figures \ref{fig:vacpw}c and \ref{fig:vacpw}d show the differential resistance as a function of bias current. For the microbridges without SiO$_x$, $dU/dI$ remains predominantly positive and increases sharply prior to failure. The coated structures exhibit maxima in the differential resistance and regions where $dU/dI < 0$, corresponding to negative differential resistance and the development of electrothermal instability. This does not, by itself, imply an exponential increase in power, but indicates strong positive feedback between Joule heating and resistance. Narrow peaks in the derivative are sensitive to the current step and the numerical differentiation method; therefore, quantitative comparison of their amplitudes requires specification of the smoothing and derivative-calculation procedures.

Plotting $dU/dI$ as a function of power, $P = UI$ (Figures \ref{fig:vacpw}e and \ref{fig:vacpw}f), allows structures of different widths to be compared on a common power scale. Without SiO$_x$, the sharp increase in differential resistance shifts from values below 1 W for narrow microbridges to values above 6 W for wider ones. In the presence of SiO$_x$, the main extrema and sign changes in $dU/dI$ occur approximately within the 0.8–3.0 W range, while individual branches extend up to \ensuremath{\approx}4.5 W. The critical point was defined as the last stable point before a sharp change in $dU/dI$, the onset of negative differential resistance, or structural failure. The maximum critical power density was \ensuremath{\approx}141 MW/cm$^2$.

\begin{figure}
    \centering
    \includegraphics[width=\linewidth]{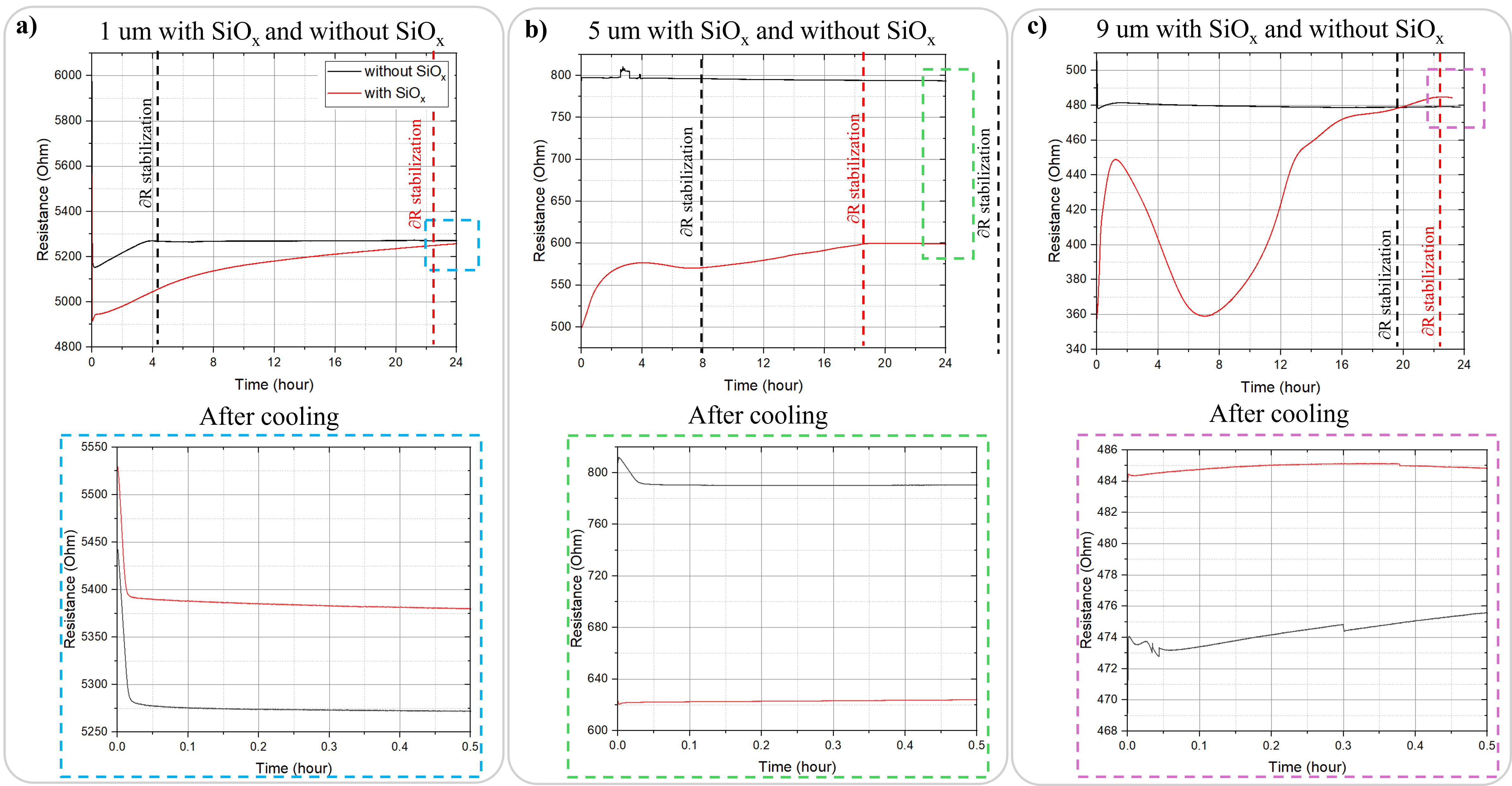}
    \caption{Change in microbridge resistance during a 24 h test at a power equal to 90\% of the critical value and upon reactivation after cooling: (a) width of 1 $\mu$m; (b) 5 $\mu$m; (c) 9 $\mu$m.}
    \label{fig:intime}
\end{figure}

To evaluate the long-term stability of the structures without SiO$_x$ and with a protective SiO$_x$ layer, the current bias was selected to correspond to a power level equal to 90\% of the critical value. The measurements were carried out over 24 h of continuous operation (Figure \ref{fig:intime}).

\FloatBarrier

For the analysis, microbridges with widths of 1, 5, and 9 $\mu$m were selected to represent the lower, middle, and upper parts of the dimensional range. On average, the resistance stabilized after \ensuremath{\approx}16 h of continuous operation and subsequently changed only slightly. The uncoated samples reached a steady-state regime faster than the SiO$_x$-coated structures (Figure \ref{fig:intime}). The initial drift may be associated with thermal annealing of the material and oxygen transport from the protective coating into the TiN$_x$ barrier layer; confirmation of this mechanism requires additional structural and chemical analysis after testing.

After the 24 h test, the electrical load was switched off and the samples were cooled to room temperature, after which the measurements were repeated for 30 min. Upon reactivation, the resistance returned to the steady-state values; the mean relative deviation for the three investigated widths was 2.17\% (lower panels of Figure \ref{fig:intime}).

This effect is presumably associated with recrystallization of the functional material as a result of substantial heating to temperatures above 450–500 °C.

\section{Conclusion}

A process for forming a three-layer TiN$_x$/Ti/TiN$_x$ material within a single vacuum cycle was developed. The TCR of the material was determined to be $\alpha = 0.00316$ °C$^{-1}$. The maximum critical power density of the microbridges reached 141 MW/cm$^2$. Under continuous operation at 90\% of the critical power, the structures remained operational for 24 h, and after cooling and reactivation, the mean deviation of the steady-state resistance was 2.17\%.

The SiO$_x$ coating substantially modifies the electrothermal characteristics: within the measurement range, the coated structures exhibited negative differential resistance and estimated temperatures of up to 730 °C at a power of approximately 2 W. Because these temperatures were obtained by extrapolating the $R(T)$ calibration beyond 250 °C, independent high-temperature thermometry is required for more accurate determination. The overall set of obtained characteristics demonstrates the potential of TiN$_x$/Ti/TiN$_x$ for use in microheaters for thermocatalytic gas sensors.
\pagebreak

\medskip
\textbf{Acknowledgements} \par 
The article was prepared with the financial support of the Ministry of Science and Higher Education of the Russian Federation (grant no. 075-15-2025-587).
Fabrication and characterization of sample were carried out at large scale facility complex for heterogeneous integration technologies and silicon+carbon nanotechnologies.





\end{document}